\documentclass[final, nomarks]{dmtcs-episciences}

\usepackage[round]{natbib}

\usepackage[utf8]{inputenc}
\usepackage{amsfonts}
\usepackage{amssymb}
\usepackage{amsmath}
\usepackage{url}
\usepackage{hyperref}
\usepackage{tikz}
\usepackage{tikz-cd}
\usepackage{subcaption}
\usepackage{underoverlap}
\usepackage{verbatim}

\newtheorem{definition}{Definition}

\author[Bill Kay et al.]{Bill Kay
  \and Audun Myers
  \and Thad Boydston
  \and Emily Ellwein\\
  \and Cameron Mackenzie
  \and Iliana Alvarez
  \and Erik Lentz}

  \title[Permutation Entropy for Signal Analysis]{Permutation Entropy for Signal Analysis}

\affiliation{
  Pacific Northwest National Laboratory, Richland, Washington, USA}

\begin{document}
\publicationdata{vol. 26:1, Permutation Patterns 2023}{2024}{9}{10.46298/dmtcs.12644}{2023-12-05; 2023-12-05; 2024-06-26}{2024-07-07}
\maketitle


\begin{abstract}
\vspace{\baselineskip} 
Shannon Entropy is the preeminent tool for measuring the level of uncertainty (and conversely, information content) in a  random variable. In the field of communications, entropy can be used to express the information content of given signals (represented as time series) by considering random variables which sample from specified subsequences. In this paper, we will discuss how an entropy variant, the \textit{permutation entropy} can be used to study and classify radio frequency signals in a noisy environment. The permutation entropy is the entropy of the random variable which samples occurrences of permutation patterns from time series given a fixed window length, making it a function of the distribution of permutation patterns. Since the permutation entropy is a function of the relative order of data, it is (global) amplitude agnostic and thus allows for comparison between signals at different scales. This article is intended to describe a permutation patterns approach to a data driven problem in radio frequency communications research, and includes a primer on all non-permutation pattern specific background. An empirical analysis of the methods herein on radio frequency data is included. No prior knowledge of signals analysis is assumed, and permutation pattern specific notation will be included. This article serves as a self-contained introduction to the relationship between permutation patterns, entropy, and signals analysis for studying radio frequency signals and includes results on a classification task.


\end{abstract}


\section{Introduction}
\label{sec:intro}
Shannon entropy was introduced in 1948 by Claude Shannon~\cite{shannon2001mathematical} to measure the uncertainty of a random variable. Specifically, given a (discrete) random variable $X$ with  support $\mathcal{X}$ and PDF $p(\cdot)$, the Shannon entropy of $X$ is given by:
\[
H(X):= -\sum_{x\in \mathcal{X}} p(x) \log_2p(x)
\]
and expresses the expected number of bits one needs to transmit to convey the outcome of $X$. Since the standard use-case for the definition of entropy is in the unit of bits, logarithms are taken to be base $2$ unless otherwise noted. Variants of Shannon entropy (e.g., Renyi entropy~\cite{bromiley2004shannon}, joint entropy~\cite{cover1999elements}) have been used to measure the information content of specific random variables in signals processing~\cite{hughes1993analysis}, quantum physics~\cite{van2021quantum},  finance~\cite{zhou2013applications} and many other application spaces. For a thorough treatment on the subject of entropy, we direct the reader to~\cite{cover1999elements}. In this document, we focus on the so-called \textit{permutation entropy}~\cite{bandt2002permutation} (PE) as applied to time series data. 

Permutation entropy  as a method extracts information from time series  by considering only the distribution of permutation patterns which occur in the data. In this way, PE demonstrates a concrete application of permutation pattern methods to real world data analysis problems. For example, PE has shown promise as a useful tool for signal processing with successful applications~\cite{amigo2012transcripts, bariviera2018analysis, cao2004detecting, morabito2012multivariate, garland2018anomaly} ranging from biomedical to financial to mechanical.
Additionally, permutation entropy has been successfully used in some Radio Frequency (RF) signal processing such as radio frequency fingerprint extraction (i.e, identifying physical radio devices from signal measurements)~\cite{deng2017radio} and detecting RF transients from their fingerprint~\cite{yuan2015detection}.
In this work we highlight the relationship between permutation patterns and RF communications modulation schemes, demonstrating the efficacy of permutation pattern-based methods on a classification problem of common RF modulations. This proof-of-concept, while a notable data science contribution in its own right, is intended to serve as a jumping off point for the use of permutation patterns in data analytics for real world RF communications use cases.
This document is intended to be self-contained;  the domain-specific vocabulary herein will be introduced before results are presented.    

The structure of the remainder of this paper is as follows: Section~\ref{sec:tspp} presents the relationships between time series and permutation patterns of interest to this study, including the PE of time series data. Section~\ref{sec:RF} serves as a primer on RF signals for the non-practitioner, describing all the preliminary subject matter needed to understand our experimental results. Section~\ref{sec:methods} describes a method for using windowed PE to study changing information content of RF signals, creating a workflow between an RF time series and windowed PE. Section~\ref{sec:results} uses the output of that workflow at multiple window sizes to perform an ML classification task, demonstrating the efficacy of permutation entropy in capturing information content of RF signals. We close with discussion in Section~\ref{sec:discussion}.


\section{Time Series and Permutation Patterns}
\label{sec:tspp}
Our principal object of study is a \textit{signal}. Using the definition from~\cite{priemer1991introductory}, a signal is ``a function that conveys information about the behavior of a system or attributes of some phenomenon.'' In this document, signals will be represented by real valued vectors (or \textit{time series}) $\mathbf{x} = (x_1, \ldots, x_t) \in \mathbb{R}^t$, where  $\{x_i\}_{i=1}^t$ are typically data points sampled at equal time intervals from some system or sensor\footnote{Notationally, many of our definitions are greatly simplified by taking the starting index of $1$, while much of the real world data we examine comes $0$-indexed. For symbolic definitions, we use $1$-indexing, but specific data analysis is displayed using $0$-indexing. The indexing used will change as needed, but will be made clear from context whenever possible.}.   Frequently, the (signed) magnitude of the values of $\mathbf{x}$ (i.e., \textit{amplitudes}) are normalized, presenting an additional preprocessing  step to make them more comparable when they are structurally similar but have very different scales. This normalization is often non-trivial due to the unknown global characteristics of the signals. PE is a function of the relative positions of the entries of $\mathbf{x}$ to one another, and is thus amplitude agnostic. Thus, PE is a function of relative patterns in a signal without suffering inefficiencies of scale or amplitdudinal noise. We make the notion of a signal pattern precise in the following definition:
  
  \begin{definition}[Vector Pattern]
  Given a real valued vector $\mathbf{x} = (x_1, \ldots, x_t) \in \mathbb{R}^t$ with distinct entries, the \textit{pattern} of $\mathbf{x}$:
  
  \[
  \pi(\mathbf{x}) := a_1 a_2 \ldots a_t,
  \]
  
  where $\{a_i\}_{i=1}^t = [t] := \{1, 2, \ldots, t\}$, and $a_i < a_j$ if and only if $x_i < x_j$.   
  \end{definition}
  
  \begin{figure*}[h]
    \centering
    \includegraphics[width=0.65\linewidth]{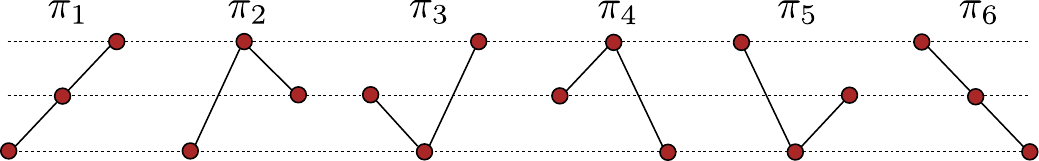}
    \caption{Example 6 possible permutation patterns for $t=3$.}
    \label{fig:perm_patterns_n3_horizonatal}
\end{figure*}
  
Informally, $\pi(\mathbf{x})$ is the permutation pattern of the symbols of $\mathbf{x}$. We remark that in signals applications, the assumption that the entries of $\mathbf{x}$ are distinct is unstated and is rather enforced by the nature of a physical system where two high precision measurements being exactly the same is unlikely. In the event that a low precision detector is used and ties occur, a sequence of observations can be made to fit the definition of a vector pattern  by breaking ties by the order in which they show up in the time series. For example, the time series $\mathbf{x} = (1.2, 3.1, -4.9)$ has permutation pattern $\pi_4$ given the 6 possible permutation patterns ($t!$ possible patterns) for signals of length $t=3$ shown in Fig.~\ref{fig:perm_patterns_n3_horizonatal}.

Given a  real valued time series $\mathbf{x}= (x_1, x_2, \ldots, x_t)$, we are frequently interested in the set of permutation patterns of specified subsequences of $\mathbf{x}$. We include some convenient notation for denoting specific subsequences of interest towards concisely stating the definition of permutation entropy. Let $\{i_1 < i_2 < \ldots <i_n\}= A \subseteq [t]$, and denote:
  
  \[
  \mathbf{x}_A := (x_{i_1}, x_{i_2}, \ldots, x_{i_n})
  \]
  the subsequence of $\mathbf{x}$ induced by $A$. Each subset $A$ of $[t]$ induces a different subsequence of $\mathbf{x}$ which in turn has a permutation pattern on the set of symbols $[n]$.

  While we can consider arbitrary subsequences of $\mathbf{x}$, permutation entropy is a function of subsequences of a fixed size $n$ (i.e., \textit{dimension}) whose entries are equally spaced by $\tau$ units (i.e., \textit{delay}). To this end, for $n, \tau\in [t]$ and for $1 \le k \leq t -(n-1) \tau $ we denote:
  
  \[
  A_{n, \tau}^k (\mathbf{x}) := (x_k, x_{k+\tau}, x_{k + 2 \tau}, \ldots, x_{k + (n-1)\tau}).
  \]
  
  Informally, $A_{n,\tau}^k(\mathbf{x})$ is the subsequence of $\mathbf{x}$ which begins at index $k$ and takes every $\tau$-th entry of $\mathbf{x}$ for $n$ steps.  For a fixed $n$, let $\{\pi_i^n\}_{i=1}^{n!}$ be an enumeration\footnote{The definitions herein are independent of the enumeration used, but in our examples we will use the lexicographic enumeration.} of the permutation patterns on the symbols $[n]$.   Denote the multiset:
  
  \[
  \Pi_{n, \tau}(\mathbf{x}) := \{\pi : \pi = \pi(\mathbf{x}_{A_{n,\tau}^k}) \text{ for } 1 \leq k \leq t-(n-1)\tau\}
 \] 
 the $n,\tau$-\textit{permutation distribution} of $\mathbf{x}$. Informally, $\Pi_{n, \tau}(\mathbf{x})$ is the collection of permutation patterns that occur in $\mathbf{x}$ by taking $n$ symbols at a time, each separated by $\tau$ spaces, with multiplicity. For fixed $n$, $\tau$, and for each $\pi_i^n$, let $p_{\mathbf{x}}^{n,\tau}(\pi_i^n)$ be the probability that a permutation drawn uniformly at random from  $\Pi_{n, \tau}(\mathbf{x})$ has pattern $\pi_i^n$.  Then we have the following:
 
 \begin{definition}[Permutation Entropy]
 Let $\mathbf{x} = (x_1, \ldots, x_t) \in \mathbb{R}^t$, and let $n$, $\tau\in [t]$. The $n$,$\tau$ \textit{permutation entropy} of $\mathbf{x}$ is given by:
 
 \[
 H_{n,\tau}(\mathbf{x}) = -\sum_{i=1}^{n!} p_{\mathbf{x}}^{n,\tau}(\pi_i^n) \log_2 p_{\mathbf{x}}^{n,\tau}(\pi_i^n).
 \]
 \end{definition}
 
Note that $H_{n,\tau}(\cdot)$ is maximized when the permutation distribution is uniform~\cite{conrad2004probability} with a value of $\log n!$. Entropy is typically a number between $0$ and $1$; to this end, we introduce the \textit{normalized} $n$,$\tau$ permutation entropy (NPE):
 \[
 h_{n,\tau}(\mathbf{x}) = -\frac{1}{\log_2 n!}\sum_{i=1}^{n!} p_{\mathbf{x}}^{n,\tau}(\pi_i^n) \log_2 p_{\mathbf{x}}^{n,\tau}(\pi_i^n).
 \]
We suppress scripts $n$ and $\tau$ when they are clear from context. In Section~\ref{sec:RF} we provide a primer on RF signals to serve as our primary use case for permutation entropic analysis.

\begin{figure}[h]
    \centering
    \begin{subfigure}[b]{.3\textwidth}
        \centering
        \includegraphics[width=.8\textwidth]{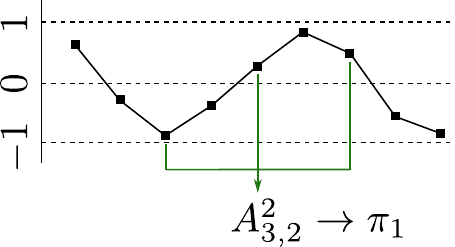}
        \caption{Time series and permutation.}
        \label{fig:example_permutation_in_signal}
    \end{subfigure}
    \begin{subfigure}[b]{.37\textwidth}
        \centering
        \includegraphics[width=.8\textwidth]{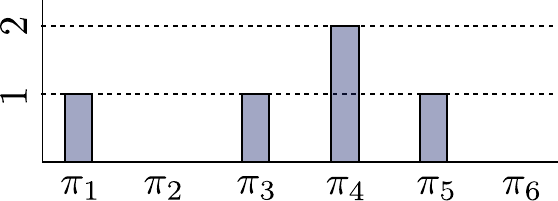}%
        \vspace{5mm}
        \caption{Histogram of permutation patterns.}
        \label{fig:historgram_of_perms_in_example}
    \end{subfigure}
    \caption{An example time series $\mathbf{x}$ and histogram of permutation patterns $\Pi_{3,2}(\mathbf{x})$. The time series is of length $t=9$ with example subsequence $A^{2}_{3,2}$ with $k=2$, $n=3$, and $\tau=2$ and corresponding permutation pattern $\pi_1$. }
    \label{fig:example_perms_of_time_series}
\end{figure}

 In Fig.~\ref{fig:example_permutation_in_signal} we show an example permutation ($k=2$) with $n=3$ and $\tau=2$ in a time series of length $t=9$. Additionally, in Fig.~\ref{fig:historgram_of_perms_in_example} we show the distribution of occurrences of permutation patterns  seen in the exemplar time series from Fig.~\ref{fig:example_perms_of_time_series} using $n=3$ and $\tau=2$. There are $5$ such permutations denoted in the histogram. Here, we have  $p(\pi_1) = 1/5$, $p(\pi_2)=0$, $p(\pi_3)=1/5$, $p(\pi_4)= 2/5$, $p(\pi_5) = 1/5$, and $p(\pi_6) = 0$. Recalling  the convention $0 \log_2 (0) = 0$  we have:
 \begin{align*}
     H_{3,2} & = -\sum_{i=1}^6 p(\pi_i) \log_2 \left ( p (\pi_i)\right)\\ &=-\frac{1}{5} \log_2 \left ( \frac{1}{5} \right )-\frac{1}{5} \log_2 \left ( \frac{1}{5} \right )-\frac{2}{5} \log_2 \left ( \frac{2}{5} \right )-\frac{1}{5} \log_2 \left ( \frac{1}{5} \right )\\
     &\approx 1.922.
 \end{align*}

The NPE can be computed as $h_{2,3} \approx 0.744$ similarly.
 
In practice, the choice of $n$ and $\tau$ is not trivial~\cite{myers2020automatic, staniek2007parameter}. One remedy is to instead use\textit{ Multi-Scale Permutation Entropy} (MSPE) in which multiple combinations of $\tau$ and $n$ are used to compute a matrix of permutation entropy values.
For example if we use the same signal as in Fig.~\ref{fig:example_permutation_in_signal} but instead of a single $n=3$ and $\tau=2$ we choose $n \in [2, 3]$ and $\tau \in [1, 2]$ we get the MSPE matrix
\begin{equation}
    H_{\rm MS} = 
    \begin{bmatrix}
    H_{2,1} & H_{2,2}\\
    H_{3,1} & H_{3,2}
    \end{bmatrix} \approx 
        \begin{bmatrix}
        0.95 & 0.99\\
        1.84 & 1.92
        \end{bmatrix},
\end{equation}
where the rows are for each dimension $n$ and the columns are for each delay $\tau$. This matrix now holds additional information that would have been lost through the choice of a single dimension and delay. For the interested reader, an early application of multiscale entropy can be found in~\cite{costa2002multiscale} which deals with physiological time series (e.g., heart rate). An early application of MSPE can be found in~\cite{aziz2005multiscale} with the same biomedical use case.   This paper includes the first application of multi-scale permutation entropy to the classification of RF signal modulations, which will be explained in Section~\ref{sec:RF}.

\section{RF signal processing background}

\label{sec:RF}

\begin{figure}
\centering
\begin{subfigure}[b]{.45\linewidth}
\includegraphics[width=\linewidth]{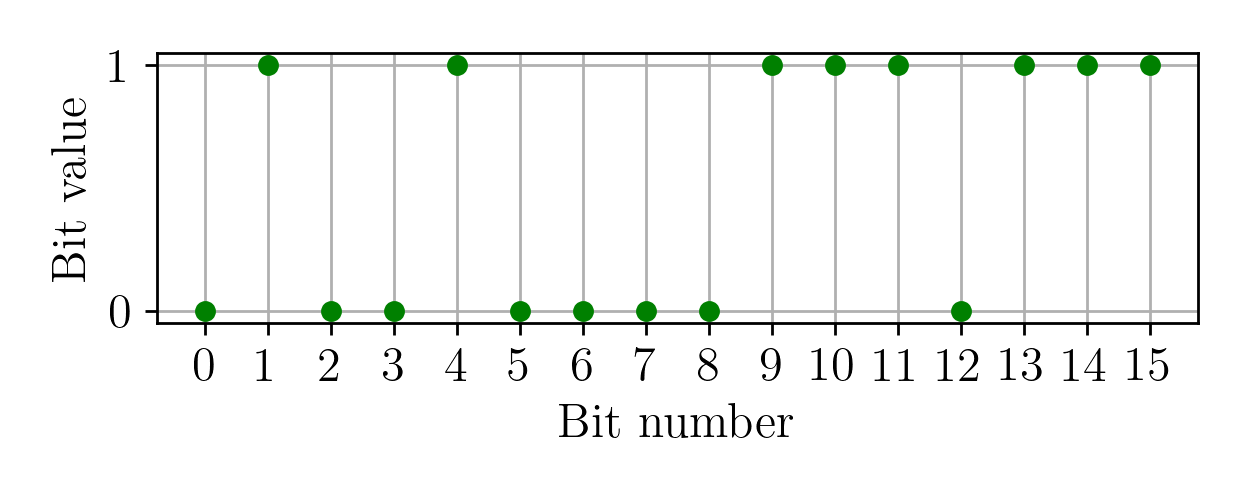}
\caption{A binary signal}\label{fig:bin}
\end{subfigure}
\begin{subfigure}[b]{.45\linewidth}
\includegraphics[width=\linewidth]{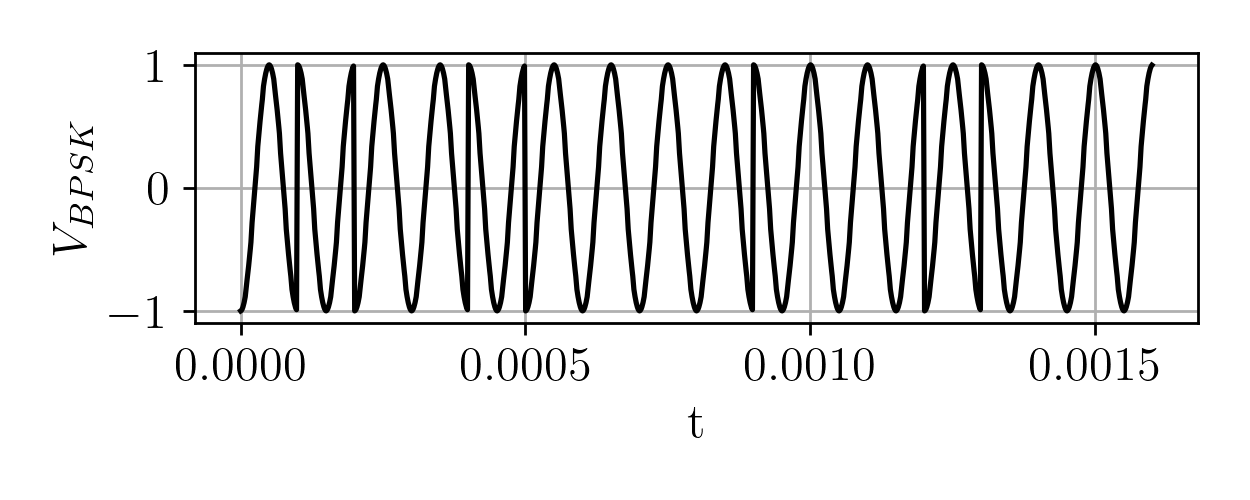}
\caption{Modulated carrier signal}\label{fig:modsig}
\end{subfigure}

\begin{subfigure}[b]{.45\linewidth}
\includegraphics[width=\linewidth]{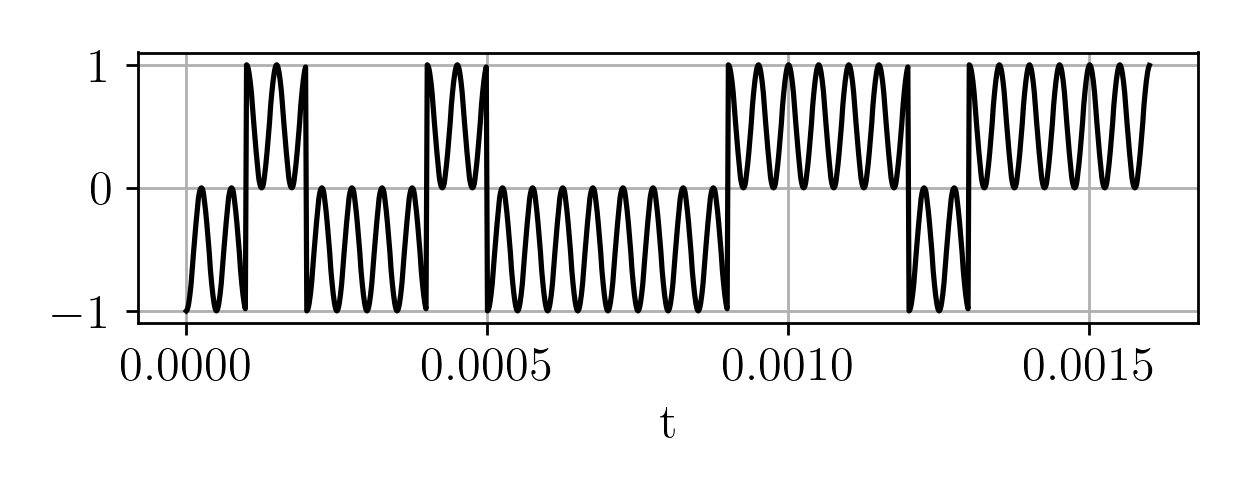}
\caption{Demodulated carrier signal}\label{fig:demodsig}
\end{subfigure}
\begin{subfigure}[b]{.45\linewidth}
\includegraphics[width=\linewidth]{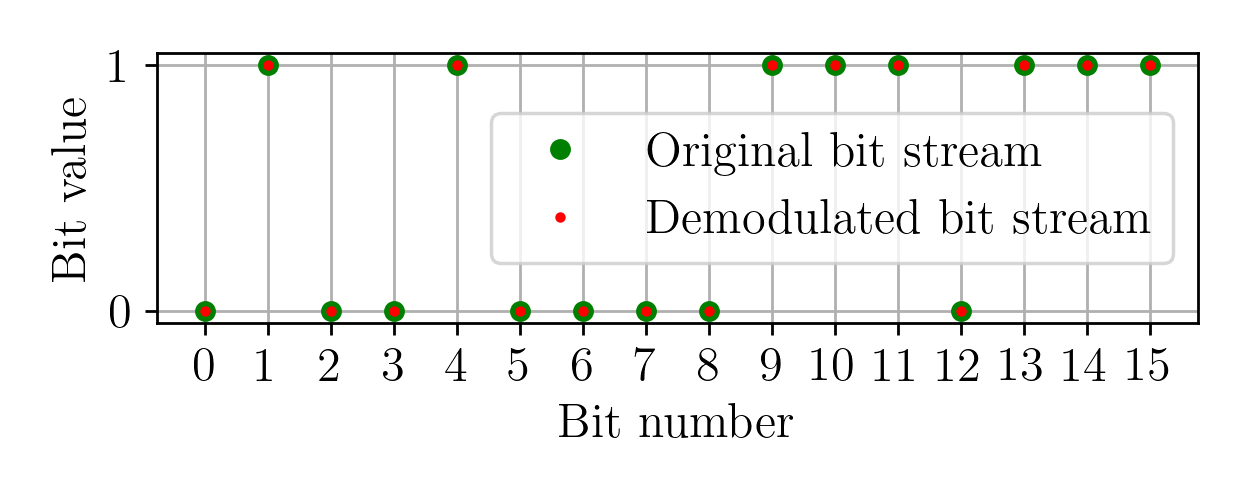}
\caption{Reconstructed binary signal}\label{fig:guess}
\end{subfigure}
\caption{An example to illustrate Binary Phase Shift Key modulation. Figure~{$(a)$} represents an input binary signal. Figure~{$(b)$} shows the input signal with the bits of the signal encoded by the phase of a carrier signal. Figure~{$(c)$} shows the output of a demodulator, which interprets the phases of the modulated signal as positive or negative. Figure~{$(d)$} reconstructs the original binary signal from the demodulated signal.}
\label{fig:modem}
\end{figure}

Radio frequency (RF) waves  model the oscillation of various media including electrical, electromagnetic, or mechanical systems in the range of $30$kHz to about $300$GHz~\cite{scarpati2021radio}. Familiar applications of RF signals include AM/FM radio, television, bluetooth, and internet communications. 
RF digital communications are generated by modulating a binary sequence into a wave-form that can be transmitted and received (see Fig.~\ref{fig:modem} for an example) to relay the desired information.
We make this specification because this use-case (e.g., sequences of \textit{bits}) is pervasive, but also because entropy is a measurement of the communication of information, and thus entropic quantification is a natural method for studying RF signals.
We will not go into details about specific frequencies or transmission methods-- an introduction on RF signals can be found here~\cite{tse2005fundamentals}. Rather, we provide a high level explanation of the process for sending binary signals via RF signals to inform our experiments. 

Broadcasting binary sequences via RF signals has two principal components. The first turns a binary sequence into a wave form (via a \textit{modulation} scheme), and the second turns a modulated wave form into a binary sequence (via \textit{demodulation}). The process of modulation/demodulation is commonplace (for example, the familiar ``modem'' is a portmanteu of ``modulator/demodulator'').  Here we describe at a high level how the commonly used binary phase shift key (BPSK) modulator/demodulator works.

BPSK modulators take as input a binary sequence and a carrier signal. Instances of the carrier signal in two different phases are concatenated, with one phase representing $1$ bits and another representing $0$ bits. The modulated signal is transmitted, and the received signal (which is a close approximation of the transmitted signal) is turned into a signal which is positive during periods meant to represent a $1$, and negative otherwise. The recovered signal is translated back to a binary string accordingly. For details about the implementation of BPSK, see~\cite{sklar2021digital}. We have included a simple example in Figure~\ref{fig:modem}.

More sophisticated modulation/demodulation schemes will be introduced in Section~\ref{sec:results}, where we show that permutation entropy serves as a high fidelity input to a convolution neural network which classifies signals by modulation mode.


\section{Methods}
\label{sec:methods}

The PE of a time series $\mathbf{x}$ is a single number which is a measure of the (lack of) information content of $\mathbf{x}$. For example, the permutation distribution $\Pi(\mathbf{x})$ is uniform if and only if $h(\mathbf{x})=1$. In the context of PE, this fact can be reasonably interpreted as the statement ``revealing the permutation pattern at any given subsequence is minimally informative'' in the sense that each pattern is equally likely to occur.  In many use cases,  it is informative to study how the PE evolves over time. For example, suppose someone is monitoring a frequency band which is expected to be random noise. The PE should be relatively high as the set of permutation patterns that occur should be close to uniformly distributed. However, a sudden drop in the PE could be indicative that a structured signal is being broadcast at this frequency. More generally, a change in PE could indicate a change in information content of a signal, and causes of these changes are frequently of interest to signal analysis practitioners. Any subsequence of $\mathbf{x}$ has coupled with it a PE, and so choosing collections of subsequences of $\mathbf{x}$ whose PE is informative is an open research area. Here, we focus on a standard method for generating a time evolving sequence of PEs of $\mathbf{x}$ by considering a collection of intervals of $\mathbf{x}$ called \textit{windows}. Similarly to this, it can be more informative to instead look at an MSPE matrix of entropy values over various dimensions $n$ and delays $\tau$. This multi-scale approach can capture more information about at what dimension and time scale the important changes in information are happening at as well as different types of changes in the signal (e.g., a change in the signal's modulation method).

Given $\mathbf{x} = (x_1, x_2, \ldots, x_t)$, an integer $1 \leq k \leq t$, and a proportion $0 \leq \alpha \leq 1$, define recursively the starting index:
\[
s_1 = 1; s_j = s_{j-1}+\lceil(1-\alpha) k\rceil \text{ for } 2 \leq j \leq t-k
\]
where $\lceil \cdot \rceil$ denotes the smallest integer \textit{strictly} greater than the input real value\footnote{For our purposes, if $(1-\alpha)k$ is an integer, $\lceil (1-\alpha)k\rceil = (1-\alpha)k +1$.}. Then the $j$th $k$-window with $\alpha$ overlap is defined:

\[
w^j_{k,\alpha}(\mathbf{x}) := (x_{s_j}, x_{{s_j}+1}, \ldots, x_{{s_j}+(k-1)})
\]

Informally, the windows are intervals of length $k$ which overlap by an $\alpha$ proportion. An example is included in  Figure~\ref{fig:window}.
\begin{figure}[h]
\[
\mathbf{x}= (\UOLunderbrace{x_1, x_{2},x_3,}[ x_{4},x_5 ]_{w^1_{5,.4}(\mathbf{x})}\UOLoverbrace{  ,x_6,}[x_7,x_8]^{w^2_{5,.4}(\mathbf{x})}\UOLunderbrace{ ,x_9,}[x_{10},x_{11}]_{w^3_{5,.4}(\mathbf{x})}\UOLoverbrace{, x_{12},x_{13}, x_{14}}^{w^4_{5,.4}(\mathbf{x})} )
\]
\caption{A length $t=14$ time series and length $k=5$ windows with overlap proportion $\alpha = .4$. Since $\alpha k = 2$, consecutive windows overlap by two.}
\label{fig:window}
\end{figure}

For a time series $\mathbf{x}$, a fixed dimension $n$, delay $\tau$, window size $k$ and overlap proportion $\alpha$ we can compute the NPE time series $(h_1, \ldots, h_\ell)$ (called the $(n,\tau, k, \alpha)$-\textit{permutation entropy profile} of $\mathbf{x}$, denoted $h_{n,\tau, k, \alpha}(\mathbf{x})$), where:

\[h_i := h_{n,\tau}(w^i_{k,\alpha}(\mathbf{x})).
\]
Here, $\ell$ is the largest integer such that $s_\ell +(k-1) \leq t$. Informally, we window $\mathbf{x}$ and compute the permutation entropy for each window. In this setup $n$, $\tau$, $k$, and $\alpha$ are tuneable parameters.  The parameters $n$ and $\tau$ are frequently selected according to heuristics or are informed by domain knowledge. A collection of automatic parameter selection tools for selecting $n$ and $\tau$ can be found in~\cite{myers2020automatic}. Suitable choices for $k$ and $\alpha$ vary based on use cases and available computation. In general, if $k$ is too small then the distribution of permutation patterns is not very informative, and if $\alpha$ is too small then the sequence of permutation entropies is coarse and can vary greatly from term to term. However, the computation of the permutation distribution scales with $k$ and the number of windows scales with $\alpha$, so taking both $k$ and $\alpha$ large can be computationally expensive. 

Instead of choosing a single value for $n$, $\tau$, $k$, and $\alpha$ one can choose multiple values of these parameters to extract different permutation pattern distributions (i.e., MSPE).  In Section~\ref{sec:results}, we show empirically that the MSPE of a time series over multiple sizes of non-overlapping (e.g., $\alpha =0$) windows yields an accuracy performance increase in a classification problem over raw signal data.


\section{Data and Results}
\label{sec:results}

We introduce in this section the dataset used to evaluate our MSPE approach, give futher detail about our evaluation method, and present the results of a comparison study where modest \textit{Convolutional Neural Networks} (CNNs) were trained to classify modulation schemes when given MPSE, spectrogram, or raw waveform data as input.
The data we work with is the Panoradio data set introduced in~\cite{scholl2019classification}  and sourced at~\cite{panoradio}.\footnote{The purpose of this article is to show the connection between permutation patterns and signal analysis. To this end, we present only a high level overview of the Panoradio data to demonstrate efficacy of permutation entropy on a classification task. However, this data is thoroughly documented, and the interested reader is encouraged to explore the source material~\cite{scholl2019classification}~\cite{panoradio}.}
This dataset is an extension of a more standard modulation classification dataset~\cite{o2018over} which further incorporates different modes of communication and baud rates\footnote{Baud rate: modulation rate in units of symbols-per-second}. 
Specifically, there are 18 different modulation modes in the data set that are shown in Table~\ref{tab:modes} in the appendix with both the modulation method and the baud rate if applicable. Additionally, Fig.~\ref{fig:modulation_mode_signals} of the appendix provides example signals from each of the 18 modulation methods. These modes cover many of the standard modulation methods used in modern RF signal processing. Plain text was encoded for digital modes, speech and music were used for analog modes, and black and white images for fax data. To make this simulated dataset more realistic there is additive white Gaussian noise at different levels, non-coherent reception and receiver frequency mismatch, and random phase and frequency offsets.

When applying MSPE to the signals we used several non-overlapping windows of the signals. Specifically, we used the full signal, the signal split into two windows, and the signal split into 4 windows. For each of these 7 windows we applied MSPE for $n \in [3,4,5,6,7]$ and $\tau \in [1, 5, 10, 15, 20, 30, 40, 50]$. This resulted in in MSPE matrices of size $(5\times 8)$, which were flattened into vectors of length 40 and concatenated into a single vector of length 280. We then used these vectors of length 280 as input into our CNN model described below. As a first point of comparison we also used the raw signal of length 2048 as input to another CNN model. A third CNN was trained on three spectrogram representations of the signal with the first spectrogram using 8 non-overlapping windows ($\alpha = 0$) of the signal each of length 256, the second using 16 non-overlapping windows each of length 128, and the third using 32 non-overlapping windows each of length 64. All three of the spectrograms were then flattened and concatenated into a single vector as input for the CNN.

\begin{figure}[h!]
    \centering
    \begin{subfigure}[b]{.49\textwidth}
        \centering
        \includegraphics[width=.99\textwidth]{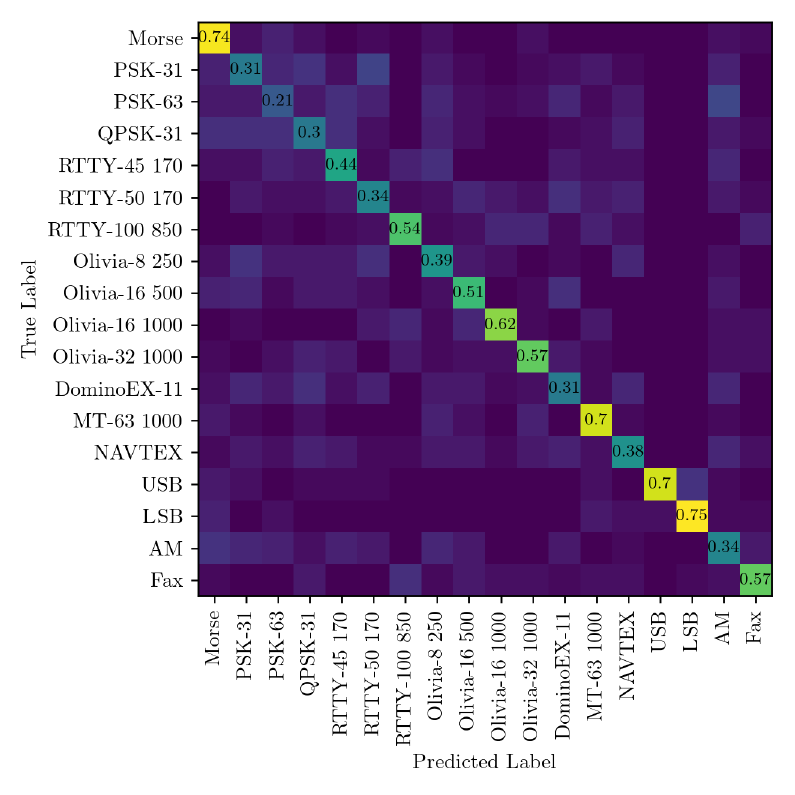}
        \caption{Confusion matrix on CNN classifier trained on signal directly with an average classification accuracy of 48.4\%.}
        \label{fig:confusion_matrix_signal}
    \end{subfigure}
    \hfill
    \begin{subfigure}[b]{.49\textwidth}
        \centering
        \includegraphics[width=.99\textwidth]{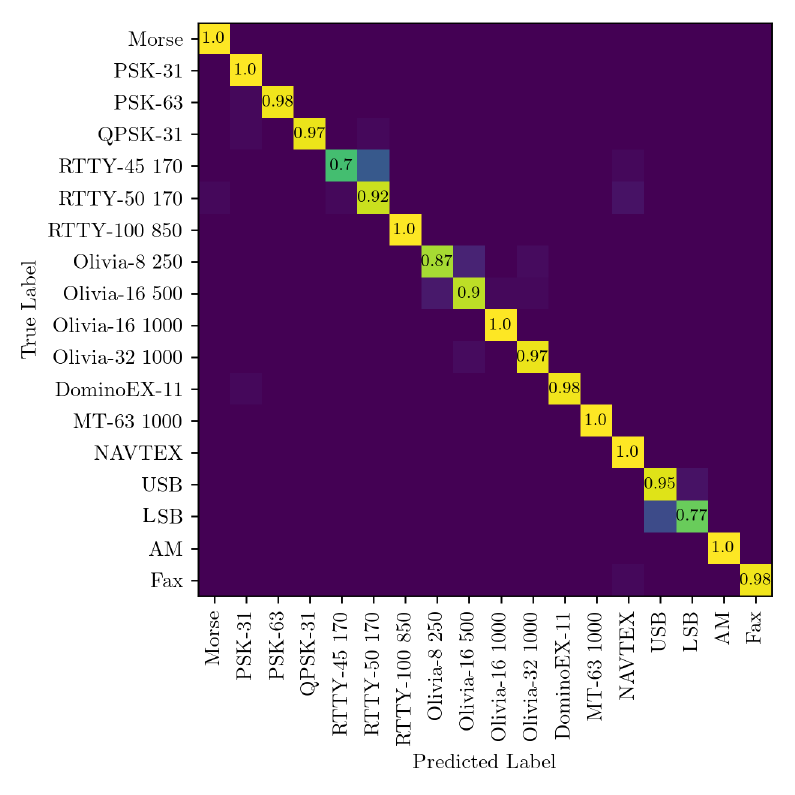}%
        \caption{Confusion matrix on CNN classifier trained on the MSPE with an average classification accuracy of 94.4\%.}
        \label{fig:confusion_matrix_MSPE}
    \end{subfigure}
    \begin{subfigure}[b]{.49\textwidth}
        \centering
        \includegraphics[width=.99\textwidth]{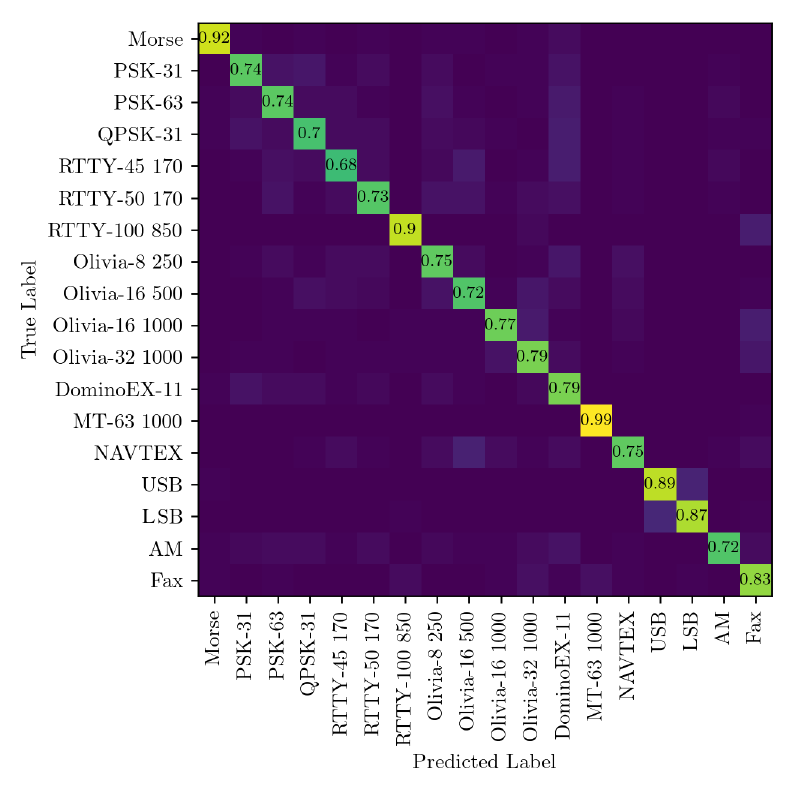}%
        \caption{Confusion matrix on CNN classifier trained on spectrograms with an average classification accuracy of 81.5\%.}
        \label{fig:confusion_matrix_MSPE}
    \end{subfigure}
    \caption{Confusion matrix for CNN trained on 25 dB raw signals (a) compared to the MSPE (b) and spectrogram results (c).}
    \label{fig:confusion_matrices}
\end{figure}

The CNNs we use have two convolution layers, each followed by maximum pooling, and one fully connected layer. We trained each CNN for 30 epochs with a batch size of 10.
The CNN trained on the raw signal, MSPE, and spectrograms had a total of 37.0K, 12.6K, and 36.5K parameters, respectively. In comparison to the original Panoradio dataset~\cite{scholl2019classification} with approximately 1.4 million parameters and 17 layers, these neural networks are significantly smaller and quicker to train.

Our first experiment emphasizes the ability for MSPE to capture the key characteristics between modulation modes in comparison to training on the raw signal. In Fig.~\ref{fig:confusion_matrices} we provide confusion matrices~\footnote{heatmaps for matrices that are indexed by true labels and predicted labels and whose $(i,j)$ entry counts the number of times label $i$ was true and label $j$ was predicted. Diagonal entries of a confusion matrix represent correct predictions} for the CNN trained on the signals directly, MSPE, and spectrograms. In summary, the MSPE CNN outperformed both the signal and spectrogram CNNs with an average test accuracy of 94.2\%, 48.4\%, and 81.5\%, respectively. In the original work classifying this dataset the authors found at a best performance of approximately 98\% using a ResNet with 41 layers and 1.4 million parameters and approximately 94\% using a standard CNN architecture that is similar but much larger than ours having 6 convolution layers in comparison to our 2 convolution layers. In comparison, our CNN trained on the MSPE vectors is much faster and comparable accurate to the classical CNN of Scholl. For the raw signal input, we believe the smaller size of our CNN is the main cause for the significant decrease in classification accuracy (48.4\%) in comparison to Scholl's classical CNN.

In Fig.~\ref{fig:confusion_matrix_MSPE} it can be seen that the MSPE CNN does not perform as well when classifying between RTTY-45 170 and RTTY-50 170 as well as USB and LSB signals. We hypothesize the difficulty between RTTY-45 and RTTY-50 is due to only the slight differences in the signals being a baud rate difference of 5. Additionally, the difficulty in classifying between USB and LSB signals is due to their similarity in signal structure with both having close to stochastic behavior with permutations not being able to clearly detect differences between noise models. However, In comparison to the Res Net from Scholl, the MSPE vectors are better at classifiying between PSK and QPSK modulation with their QPSK classification being 86\% and ours at 97\%. We believe thi is due to MSPE being able to measure the increase signal complexity from PSK to QPSK.

\begin{figure*}[h]
    \centering
  \includegraphics[width=0.8\linewidth]{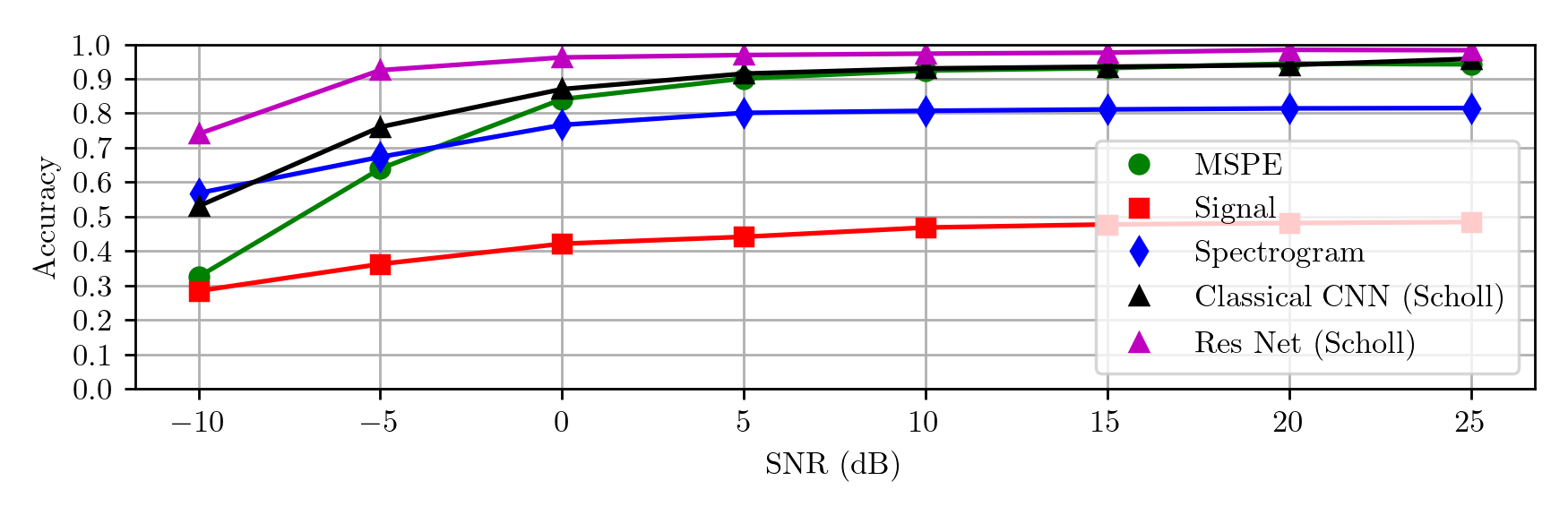}
    \caption{Average classification accuracy over SNRs of CNN trained on MSPE (green circles), the raw signal (red squares), and spectrograms (blue diamonds). Results from Scholl are shown for comparison: classical CNN trained (black triangles) and Res Net (purple tringles).}
    \label{fig:accuracy_vs_SNR}
\end{figure*}

Our second experiment is to determine the ability for the CNNs to operate when additive white Gaussian noise is increased in the signal by decreasing the Signal to Noise Ratio (SNR). The SNR is calculated as $${\rm SNR} = 20\log_{10} {\left( \frac{A_{\rm signal}}{A_{\rm noise}} \right) },$$ where $A$ for signal and noise are calculated as the root mean square of the noise-free signal and  just the additive noise, respectively. In Fig.~\ref{fig:accuracy_vs_SNR} we show the classification accuracy of each CNN for each level of noise with SNRs ranging from -10 to 25 dB.

As shown, the best performing CNN's are from Scholl, but these are significantly larger and take correspondingly longer to train. However, above 0 dB SNR both the MSPE and classical CNN perform comparably.
As a more fair comparison using just our CNN, we can see that when trained on the raw signal the accuracy is 48.4\% at 25 dB and then degrades slowly  with decreasing logarithmic SNR, ending in an accuracy of 28.4\%. In comparison, the MSPE CNN had above 90\% classification accuracy down to an SNR of 5 dB and then quickly decreasing to 32.1\% accuracy at -10 dB. Lastly, the CNN trained on the spectrograms performed moderately well with a classification accuracy above 80\% accuracy above 5 dB SNR and then only dropping to 57.8\% accuracy at -10 dB SNR. These results showing the ability for MSPE to outperform other data representations of the signals when using a comparable CNN suggests that MSPE is a suitable tool for RF signal processing.


\section{Discussion}
\label{sec:discussion}
In this article we discussed a connection between permutation patterns and the information content of signals. We demonstrated that information content for RF signals can be captured by the distribution of permutation patterns embedded in time series representations of the signals by showing that a classification problem displayed a marked accuracy improvement when considering only multiple-scale permutation entropy over the entire signal. Our simple CNN with 12.6K parameters trained on MSPE even outperformed a CNN on 1.4 million parameters and a $41$ layer ResNet with 1.4 million parameters  both trained on raw signal data. 

While the modulation classification performance serves as a proof-of-concept that the distribution of permutation patterns which occur within time series signal data captures critical information about RF signals (with simple MSPE models beating large  CNN and ResNet models on raw data), modulation dependent methods face limitations in practice. For example, the 18 modulations in the Panoradio data set are in no way comprehensive; many use cases come with bespoke modulation schemes and adversarial methods can be used to confound modulation detection schema~\cite{lin2020adversarial} ~\cite{hameed2020best}. However, minor signal perturbations should not drastically affect the distribution of permutation patterns within a signal. A qualitative analysis of which properties of a signal that MSPE methods capture is the first step towards modulation agnostic signal information analysis which is an active and ongoing research area.
\section{Future Permutation Patterns Work}
\label{sec:future}

While our application herein has been RF specific, we remark here that our general approach of \textit{applied permutation patterns} gives rise to a set of tools with respect to pattern statistics in general time series, and need not be focused on a specific use case. To this end, we  lay out a collection of questions and future directions in the more general regime to foster both theoretical work and follow on applications. Broadly, given a time series $\mathbf{x}$ and choices of $n$, $\tau$, $k$, and $\alpha$, we ask questions about the set $\Pi_{n,\tau}(\mathbf{x})$, the parameter $h_{n,\tau}(\mathbf{x})$, and the statistic $h_{n,\tau,k,\alpha}(\mathbf{x})$ (defined in Section~\ref{sec:tspp} and Section~\ref{sec:methods}). 

Recall that Shannon Entropy is uniquely maximized by the uniform distribution, and thus $h_{n,\tau}(\mathbf{x}) = 1$ if and only if $\Pi_{n,\tau}(\mathbf{x})$ is distributed uniformly as a multi-set. To this end, the set of pairs $\{(n_i, \tau_j)\}$ for which $h_{n_i,\tau_j}(\mathbf{x}) =1$ provides a measure for \textit{how} uniform a time series is. Conversely, the set of symbols $\{(n_i, \tau_j)\}$ for which $h_{n_i,\tau_j}(\mathbf{x}) \neq 1$ is a measure of the scales on which a time series is informative. To this end, we ask the following two questions:
\begin{itemize}
\item \underline{Question 1}: For which values of $n$ and $\tau$ is $h_{n,\tau}(\mathbf{x}) = 1$?

\item\underline{Question 2}: For which values of $n$ and $\tau$ is $h_{n,\tau}(\mathbf{x}) \neq 1$? For which values of $n$ and $\tau$ is $h_{n,\tau}(\mathbf{x})$ minimized?
\end{itemize}

We focus here on statistics of the multi-scale permutation entropy of a time series. For a collection of choices of parameters $\{(n_i, \tau_j, k_s, \alpha_t)\}$, $\{h_{n_i, \tau_j}(\mathbf{x})\}$ is a collection of empirical observations of  distributions $\Pi_{n_i, \tau_j}(\mathbf{x})$, and further $\{h_{n_i, \tau_j, k_s, \alpha_t}(\mathbf{x})\}$ is a set of \textit{sequences} of observations of distributions. We ask the following two questions:
\begin{itemize}

\item \underline{Question 3}: Can one provide an approximation of a \textit{maximum likelihood estimator} $\theta$ given the above sets of permutation entropy observations of various signal types (jointly and independently)?

\item \underline{Question 4}: Given a uniform prior distribution, can one find the posterior distribution of permutation patterns of $\mathbf{x}$?
\end{itemize}

Combinatorially, we find that the \textit{converse} question of time series which are realizable given a distribution of permutation patterns has connections to covering, packing, and universal sequences~\cite{godbole2018threshold}. To this end, we ask:

\begin{itemize}
\item \underline{Question 5}: Given a permutation pattern distribution $\Pi$, for which sequences $\mathbf{x}$ do there exist parameters $n$ and $\tau$ such that $\Pi_{n,\tau} (\mathbf{x}) = \Pi$? Which sequences are minimal in length with this property?

\item \underline{Question 6}: Given a real value $h$ (resp. sequence of real values $\mathbf{h})$, for which sequences $\mathbf{x}$ do there exist parameters $n$, $\tau$ (resp. $n$, $\tau$, $k$, $\alpha$) for which the normalized permutation entropy $h_{n,\tau}(\mathbf{x})=h$ (resp. normalized permutation entropy profile $h_{n, \tau, k, \alpha}(\mathbf{x}) = \mathbf{h}$. Which sequences are minimal in length with this property?

\end{itemize}
\acknowledgements
\label{sec:ack}

The research described in this paper was conducted under the Laboratory Directed Research and Development Program at Pacific Northwest National Laboratory, a multi-program national laboratory operated by Battelle for the U.S. Department of Energy.

\nocite{*}
\bibliographystyle{abbrvnat}
\bibliography{thebib}
\label{sec:biblio}


\newpage

\section{Appendix}

\begin{table}[h!]
\centering
\caption{Modes in Panoradio dataset~\cite{panoradio} with corresponding modulation method and baud rate.}
\label{tab:modes}
\begin{tabular}{lll}
\textbf{Mode}           & \textbf{Modulation} & \textbf{Baud Rate} \\ \hline
Morse                   & OOK                 & Variable           \\
PSK31                   & PSK                 & 31                 \\
PSK63                   & PSK                 & 63                 \\
QPSK31                  & QPSK                & 31                 \\
RTTY 45/170             & FSK, 170 Hz shift   & 45                 \\
RTTY 50/170             & FSK, 170 Hz shift   & 50                 \\
RTTY 100/850            & FSK, 850 Hz shift   & 850                \\
Olivia 8/250            & 8-MFSK              & 31                 \\
Olivia 16/500           & 16-MFSK             & 31                 \\
Olivia 16/1000          & 16-MFSK             & 62                 \\
Olivia 32/1000          & 32-MFSK             & 31                 \\
DominoEX                & 18-MFSK             & 11                 \\
MT63 / 1000             & multi-carrier       & 10                 \\
Navtex / Sitor-B        & FSK, 170 Hz shift   & 100                \\
Single-Sideband (upper) & USB                 & -                  \\
Single-Sideband (lower) & LSB                 & -                  \\
AM broadcast            & AM                  & -                  \\
HF/radiofax             & radiofax            & -                  \\ \hline
\end{tabular}
\end{table}

\begin{figure*}[h]
    \centering
    \includegraphics[width=0.99\linewidth]{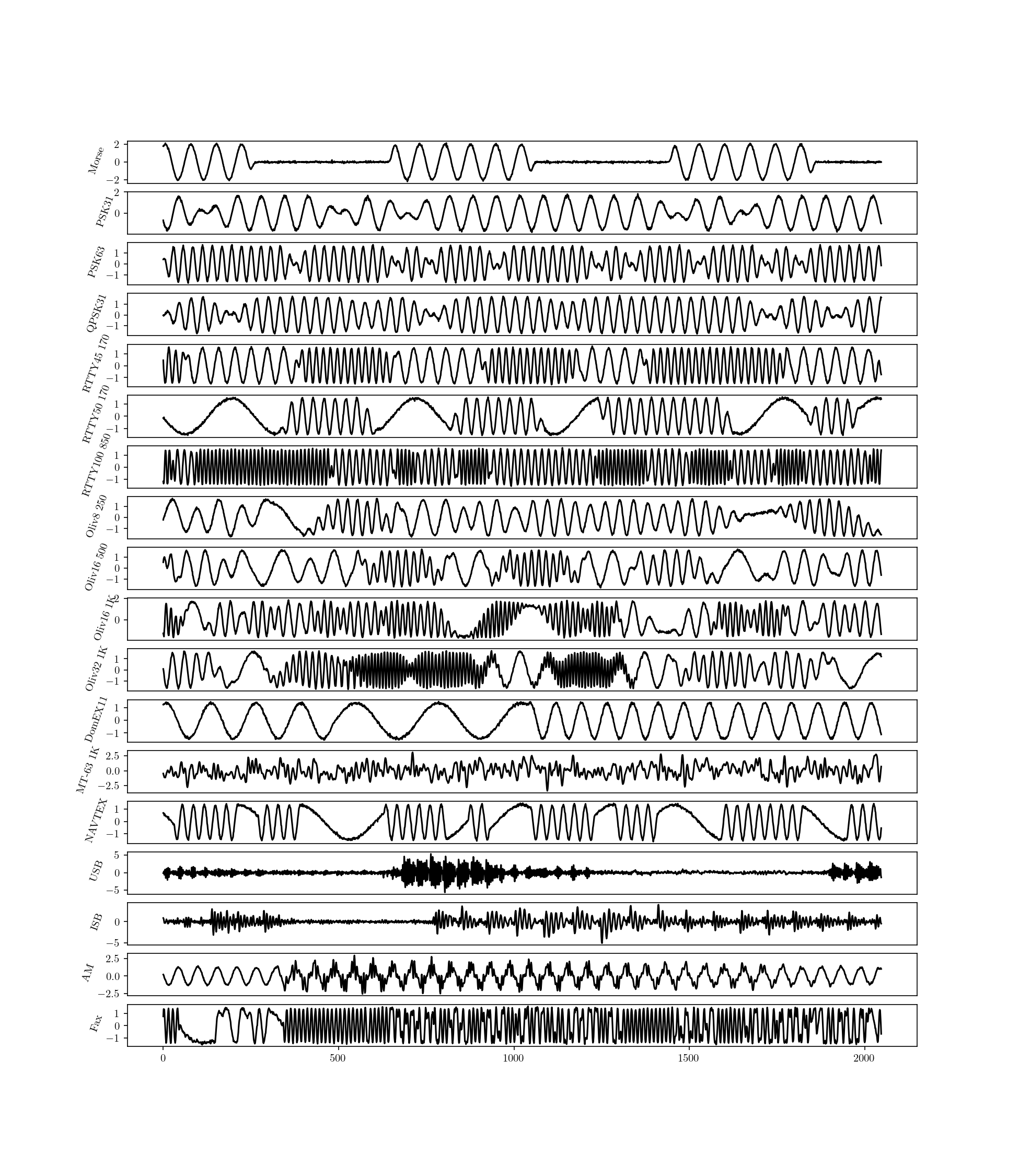}
    \caption{Example signals from each modulation modes at 25 dB SNR. The training data is generated by modulating various signals. Plain text was encoded for digital modes, speech and music were used for analog modes, and black and white images for fax data. }
    \label{fig:modulation_mode_signals}
\end{figure*}

\end{document}